\documentclass[aps,prl,showpacs,twocolumn,amsmath]{revtex4}

\usepackage{amssymb}
\usepackage{graphicx,color}
\usepackage{epsfig}
\begin{document}

\date{\today}

\newcommand{\rmd}{{\rm d}}
\newcommand{\rme}{{\rm e}}
\newcommand{\be}{\begin{equation}}
\newcommand{\ee}{\end{equation}}
\newcommand{\reff}[1]{(\ref{#1})}
\newcommand{\eref}[1]{Eq.~(\ref{#1})}
\newcommand{\erefs}[1]{Eqs.~(\ref{#1})}
\newcommand{\achtung}[1]{{\bf\textcolor{blue}{\small (#1)}}}

\title{Statistics of the Work  in Quantum Quenches, Universality\\ and 
the Critical Casimir Effect.}

\author{Andrea Gambassi$^1$ and Alessandro Silva$^2$}
\affiliation{$^1$SISSA -- International School for Advanced Studies and INFN,
  via Bonomea 265, 34136 Trieste, Italy\\
  $^2$ICTP -- International Centre for Theoretical Physics,
  P.O. Box 586, 34014 Trieste, Italy}


\begin{abstract}
We consider the dynamics of an isolated quantum many-body system after a sudden change of one control parameter, focusing on the statistics of the work done.  We establish a connection between the generating function of the distribution of the work and the partition function of a classical system in a film geometry. Using this connection, we first show that the scaling of the fidelity susceptibilities close to a quantum phase transition can be understood in terms of the critical behavior of the excess entropy and specific heat in the classical model. Remarkably, we show that the statistics of the work close to the threshold and to criticality is connected to the so-called critical Casimir free energy  which is responsible for the interaction between the boundaries of the classical system. On the basis of this relation, we highlight the emerging universal features of the statistics of the work. Our findings are exemplified for a global quench of the transverse Ising chain at zero temperature.
\end{abstract}

\maketitle

Universality, i.e., the insensitivity to microscopic details, is one of the key discoveries of modern condensed matter and statistical physics. We definitely owe to this concept the possibility to describe the low-temperature behavior of strongly correlated quantum systems in terms of simple effective models which account for just a few relevant interactions and symmetries. This is of crucial importance in many contexts, among others in  the study of the low-temperature physics of real materials, where the overwhelming number of details to be accounted for in an ab initio description would definitely hinder all progress. 

Notwithstanding the insight into universality at equilibrium, the investigation of its emergence in the dynamics of quantum many-body systems started only recently, inspired by a series of ground breaking experiments in the context of cold atoms in optical lattices \cite{Gre-Kino, Polkovnikov10}. These systems constitute a sort of ideal nonequilibrium laboratory: long coherence times and an exquisite control on the system 
made it possible to observe interesting phenomena, such as the collapse and revival cycles of a system taken abruptly across the Mott-superfluid phase boundary~\cite{Gre-Kino}. 
A common protocol employed to take the system out of equilibrium is the so-called \it quantum quench\rm, i.e., a rapid or slow change in time of one of the parameters controlling the system Hamiltonian~\cite{Polkovnikov10}. A great deal of recent research focusing on these protocols has uncovered various examples of universal features, from the scaling of  the energy density for slow quenches across quantum critical points~\cite{Zurek} to the the asymptotics of correlators after a quench at a conformally invariant quantum critical point~\cite{Calabrese}.

In the light of thermodynamics, quantum quenches are essentially thermodynamic transformations~\cite{Entropy,Silva08}, which are naturally dscribed by three standard thermodynamic quantities: change in entropy, transferred heat~\cite{Entropy}, and work $W$ done on the system~\cite{Silva08}. Focusing on the latter, it is well know both in classical and quantum physics that $W$ is in general a random variable, fluctuating in different realizations of the same protocol and described by a probability distribution $P(W)$~\cite{Jarzynski,Lutz}.   The purpose of the present study is to investigate the emergence of universal features in the statistics $P(W)$ of the work done on the system by changing abruptly one of its parameters~\cite{Silva08,work}. In order to extract all universal aspects of $P(W)$ we connect the generating function of the work distribution to the partition function of a classical system in a film geometry. 
We first use this mapping to gain insight into the scaling of the fidelity susceptibilities 
close to quantum critical points~\cite{fidelity}. Afterwards, we discuss the universality of the statistics of the work $W$ close to the threshold by uncovering its connection to the universal critical Casimir force (see, e.g., Ref.~\cite{cc-rev}) which describes the interaction between the boundaries in the classical model. In order to substantiate our general findings we focus on the example of a quench of the transverse field in the quantum Ising chain close to criticality.

Virtually all instances of quantum quenches considered so far in the literature consist of the change of a 
parameter $g$ of a many-body Hamiltonian $H(g)$ from an initial value $g_i$ to a final one $g_f$.
In this setting, the work $W$ done on this closed system can be quantified measuring the energy before and after the quench. If the quench starts in the ground state
 $| \Psi_0(g_i) \rangle$ of the initial Hamiltonian, the probability distribution of 
the work $W$ is
$P(W)=\sum_{n\ge 0}\;\!\!\delta(W\!-\![E_n(g_f)-E_0(g_i)]) 
|\langle\Psi_0(g_i)| \Psi_n(g_f) \rangle  |^2$
where $| \Psi_n(g_f) \rangle$ are the eigenstates of energy $E_n(g_f)$ of the final Hamiltonian and $P$ vanishes for 
$W<\Delta E_0 \equiv E_0(g_f)-E_0(g_i)$.
The characteristic function $G(t)\equiv \int_{-\infty}^{+\infty}\!\!\rmd W \rme^{-iWt}P(W)$ of $P(W)$ takes the form
\be
\label{Loschmidt}
G(t)=\langle \Psi_0(g_i)| \rme^{iH(g_i) t} \rme^{-iH(g_f) t} | \Psi_0(g_i)\rangle,
\ee
where here $t$  is the variable conjugate to $W$~\cite{Silva08,Lutz}. 
$|G(t)|^2$ is known as Loschmidt echo and has been studied in various contexts  ranging from quantum chaos and quantum information to strongly correlated systems~\cite{ELoschmidt}.

Essentially, the characteristic function $G(t)$ can be seen as a partition function after a proper analytic continuation to real "times" $t$. This simple, yet crucial observation yields insight into the universal features of $P(W)$. In order to see this, consider \eref{Loschmidt}, and perform an analytic continuation to imaginary "time" $t \rightarrow -i R$. We obtain
\begin{eqnarray}
\label{partition}
G(R)&=&\rme^{-\Delta E_0 \, R}\times Z(R), \quad \mbox{where}\\ \nonumber
Z(R)&=&\langle \Psi_0(g_i) | \left[\rme^{-(H(g_f)-E_0(g_f))}\right]^R | \Psi_0(g_i) \rangle.
\end{eqnarray}
For a $d$-dimensional \emph{quantum} system possessing a $d+1$-dimensional
\emph{classical} correspondent, $Z(R)$ is the partition function of the latter on a "strip"  
of thickness $R$ with two boundaries described by the boundary states $| \Psi_0(g_i) \rangle$.
If the system
has linear extension $L$ it is now possible to  separate three contributions to the free energy
$- \ln G(R)=L^d[R \times f_b  + 2 f_s + f_c(R)]$, 
where $f_b=\Delta E_0/L^d$ is the bulk free energy density and $f_s$ is 
the surface free energy associated to each of the two (identical) boundaries of the film. The remaining contribution $f_c(R)$ represents an effective interaction between these two boundaries, which generically decays to zero at large separations $R$.
In particular, one can easily identify
\be
\label{zeta}
Z(R)= 
\rme^{-L^d[2  f_s + f_c(R)]}.
\ee
Upon approaching a critical point, $f_c(R)$ becomes universal and causes the so-called critical Casimir effect~\cite{cc-rev}.

Relating quantum quenches to the statistical physics of classical confined systems makes it possible to gain insight into the scaling of the fidelity susceptibility, a quantity recently introduced for the characterization of quantum phase transitions~\cite{fidelity}. The fidelity ${\cal F} (g_i,g_f)\equiv| \langle \Psi_0(g_i) | \Psi_0(g_f)\rangle |$,
determines, inter alia, the weight of the peak of $P(W)$  at the threshold $W=\Delta E_0^+$  
Indeed, if the quantum system has a critical point at $g=g_c$, the generalized  susceptibilities 
$\chi_n(g_i,g_f)\equiv -L^{-d} \partial^n_{g_f} \ln {\cal F}(g_i,g_f)$
develop non-analytic behaviors~\cite{fidelity} and in particular the 
so-called \emph{fidelity susceptibility} $\chi_2$ scales as 
\be
\label{scaling}
\chi_2(g,g) \simeq | g - g_c |^{\nu d -2}. 
\ee
This scaling can be straightforwardly understood in terms of the statistical mechanics of the corresponding $d+1$-dimensional classical systems with boundaries, for which $g$ acts as a temperature-like variable controlling the distance from the bulk critical point. In fact,  by inserting  in \eref{partition} a resolution of the identity in terms of the eigenstates of $H(g_f)$ and using \eref{zeta} for $R \rightarrow +\infty$ one can express the fidelity  ${\cal F}$ in terms of the surface
free energy $f_s$ as 
${\cal F}= \rme^{-L^d f_s}$ and therefore $\chi_n(g_i,g_f)= \partial^n_{g_f} f_s(g_i,g_f)$.
This expression clearly reveals that $\chi_1$ is the so-called \emph{excess internal energy} 
while the fidelity susceptibility $\chi_2$ is the \emph{excess specific heat} \cite{diehl} of the $d+1$-dimensional classical confined system.
For the latter, it is known~\cite{diehl} that $\chi_2 \simeq | g - g_c |^{-\alpha_s}$ close to $g_c$, 
where $\alpha_s$ is the
excess specific heat exponent given by
$\alpha_s=\alpha+\nu$ \cite{diehl} where $\alpha$ and $\nu$ are the corresponding standard bulk critical exponents of the classical $d+1$-dimensional system. They satisfy the
hyperscaling relation $\alpha+\nu (d+1) =2$ which yields immediately the scaling in \eref{scaling}. 
In passing, we mention that mixed derivatives of $\ln {\mathcal F}$ with respect to $g_f$ and $g_i$ are also expected to be characterized by critical exponents which generally differ from $\alpha_s$  even though, in most of the cases, they can be expressed in terms of known bulk exponents~\cite{diehl}.

Setting forth an argument based on the physical analogy between critical phenomena in confined systems and quantum quenches, one can now show that $P(W\gtrsim \Delta E_0)$ 
close to the threshold $\Delta E_0$ \emph{and} to criticality $g\simeq g_c$ is \emph{universal} and determined by the asymptotic expression of the critical Casimir free energy $f_c$.  
Close to a critical point, $f_c$ takes the scaling form 
\be
\label{scalingfunction}
f_c(R) = R^{-d} F(R/\xi_+),
\ee
where $\xi_+$ coincides with the exponential correlation length of the system in the bulk disordered phase and $F(x)$ is a \emph{universal scaling function} which depends only on the surface and bulk universality classes and not on the detailed values of $g_i$, $g_f$, etc. \cite{cc-rev}. While the large-$R$ decay of $f_c$ is $\propto R^{-d}$ at criticality, away from the critical point it is dictated by the asymptotic expansion of $F(x\!\gg\!1)$. Generically, it takes the form 
\begin{eqnarray}\label{asympt}
F(x) = A(x)\;\rme^{-qx}+ \dots,
\end{eqnarray}
where the power law $A(x)$, $q\in {\mathbb R}^+$, as well as $F$, depend, inter alia, on being above or below the bulk critical point.  For the quantum (classical) Ising model in $d\!=\!1$ ($2$) $q\in \{1,2\}$ \cite{Has-10},
where the correlation length $\xi_+ \!\equiv\! a|g-g_c|^{-\nu} $ is identified with the inverse of the mass $m$ of the lightest particle of the quantum model in the paramagnetic phase. 
Performing the analytic continuation $R\mapsto it$ we obtain for $G(t)$: 
\begin{eqnarray}\label{expansiongenerating}
G(t) = \rme^{-i\Delta E_0 t -L \times 2 f_s} \;\left(1+ A(imt)\;\rme^{-iqmt}+ \dots \right), 
\end{eqnarray}
which immediately shows that $P(W)$ consists of a delta-function peak at 
$w\equiv W -\Delta E_0=0$ and of a 
continuum for $w\!>\! qm$, where $P(W)$ close to the threshold is determined by $A(x)$. It is important to stress  at this point that while one would be naively tempted to associate the small-$w$ behavior of $P(W)$ to the large-$t$ expansion of the generating function $G(t)$ -- which does not necessarily coincide with the one on the imaginary axis \eref{asympt} -- the correct procedure, instead, is the one described above and discussed in more detail on the example given below.

It is now useful to illustrate the ideas presented above in the specific case of the
quantum Ising chain subject to a quench of the transverse field~\cite{Sachdev}. The  Hamiltonian is
\be
H(g)=-J\,\mbox{$\sum_i$} \left(\sigma^{x}_i \sigma^{x}_{i+1}+g
\sigma_i^z\right),
\label{Ising}
\ee
where $\sigma^{x,z}_i$ are the spin operators at lattice site $i$ (with lattice spacing $a\!=\!1$), 
$J$ ($=\!1$  in what follows) is an overall energy scale, and  $g$ is the strength of
the transverse field. The one-dimensional quantum Ising model is the
prototypical, exactly solvable example of a quantum phase
transition~\cite{Sachdev}, with a quantum critical point at $g_c\!=1\!$ separating 
a quantum paramagnetic phase at $g\!>\!g_c$
from a ferromagnetic one at $g\!<\!g_c$. $H(g)$ can be easily analyzed 
after a Jordan-Wigner transformation~\cite{Sachdev} in terms of fermionic operators $c_k$ of momentum
$k$:  $H=2 \sum_{k>0}\;\hat{\Psi}^{\dagger}_k\; \hat{H}_k\;\hat{\Psi}_k$, where
$\hat{\Psi}_k=(c_k ,c^{\dagger}_{-k} )^T, $ and 
$\hat{H}_k=(g-\cos k )\hat{\sigma}_z-(\sin k)\hat{\sigma}_y$. This $2\times2$ matrix can be diagonalized 
with a rotation $U(\theta) \equiv \exp(-i\theta\hat\sigma_x)$ of an angle $\theta_k=1/2\arctan[(\sin k)/(g-\cos k)]$ which gives
$U^\dagger(\theta_k)\hat{H}_k U(\theta_k) = {\rm sign}(g\!-\!\cos k)E_k \hat{\sigma}_z$,
where $E_k= [(g\!-\!\cos k)^2\!+\!\sin^2k]^{1/2}$.
Accordingly, $H$ takes the form
\begin{eqnarray}
H=\sum_{k>0}\;E_k\;(\gamma^{\dagger}_k\gamma_k-\gamma_{-k}\gamma^{\dagger}_{-k}),
\end{eqnarray}
where we have conveniently introduced the quasi-particles $\gamma_k$'s in terms of the $c_k$'s:  
\be
\hat{\gamma}_k= (\gamma_k,\gamma^{\dagger}_{-k} )^T =
       \rme^{i\delta_k} U^\dagger(\theta_k+\delta_k) \hat{\Psi}_k,
\label{eq:rotation}
\ee
with $\delta_k=0$ for $g-\cos k>0$ and $\delta_k=\pi/2$ 
otherwise.

The statistics $P(W)$ of the work $W$ done during
a quench of the transverse field $g$ from $g_i$ to $g_f$ can be  calculated by expressing
$|\Psi_0(g_i)\rangle$ 
in terms of the eigenstates $|\Psi_n(g_f)\rangle$ of $H(g_f)$.
This can be done by using \eref{eq:rotation} which yields the relation 
$\hat{\gamma}_k(g_f)=\rme^{i\Delta\delta_k}\;U^{\dagger}( \Delta \alpha_k)\;\hat{\gamma}_k(g_i)$
between the operators $\hat{\gamma}_k(g_i)$ and $\hat{\gamma}_k(g_f)$ which allow the diagonalization of 
$H(g_i)$  and  $H(g_f)$, respectively, where $\Delta \delta_k=\delta_k(g_f)-\delta_k(g_i)$ and $\Delta \alpha_k=\theta_k(g_f)-\theta_k(g_i)+ \Delta\delta_k$.
Since by definition
$\gamma_k(g_i) | \Psi_0(g_i) \rangle \,\,=\,\, \rme^{-i\Delta \delta_k}\;\left[ \cos(\Delta \alpha_k)\gamma_k(g_f)-i\sin( \Delta \alpha_k)\gamma_{-k}^\dagger(g_f) \right] | \Psi_0(g_i) \rangle=0$, one can finally express $| \Psi_0(g_i) \rangle$ as
\begin{eqnarray}\label{state}
| \Psi_0(g_i) \rangle \!\!&=&\!\! {\cal N}\rme^{\sum_{k>0} i\tan(\Delta \alpha_k)\gamma^{\dagger}_k(g_f)\gamma^{\dagger}_{-k}(g_f)}
| \Psi_0(g_f)\rangle
\end{eqnarray}
where
${\cal N}=\exp[ -\frac{1}{2}\;\sum_{k>0}\;\ln\left( 1+\tan^2\Delta \alpha_k  \right)  ]$
is a normalization constant~\cite{note2} and the sum runs over the momenta in $[0,\pi]$ allowed by the finite extent of the system. Inserting now \eref{state} into \eref{Loschmidt} one finds
\be
\label{result1}
\ln G(t)= -i \Delta E_0 t + \sum_{k>0}\;\ln\left[ \frac{1+(\tan \Delta \alpha_k)^2 \rme^{-2iE_k(g_f)t} }{1+ \tan^2\Delta \alpha_k}\right].
\ee

According to \eref{state} the fidelity is given by
${\cal F}(g_i,g_f)\!\!= \!\! |{\cal N}| \equiv  \exp[-L f_s(g_i,g_f)]$,
where $f_s(g_i,g_f)=(2L)^{-1} \sum_{k>0}\;\ln\left[ 1+\tan^2\Delta \alpha_k \right]$  is nothing but the surface free energy of the classical two-dimensional (semi-infinite) Ising model. The successive derivatives with respect to $g_f$ of this expression give the first two
susceptibilities, analogous to the excess entropy and specific heat, e.g.,
\be
\label{chi1}
\chi_1=\partial_{g_f} f = \frac{1}{L} \sum_{k>0} \tan(\Delta \alpha_k) \partial_{g_f} \theta_k({g_f}). 
\ee
Close to criticality these sums are dominated by low-momentum modes and  therefore they can be evaluated in the scaling limit from the outset~\cite{note}, which yields  
\be
\label{angleappr}
\theta_k(g\gtrless 1) \simeq \pm (1/2)\arctan[k/m(g)],
\ee
for $\theta_k$ at small $k$, where $m(g)=|g-1|$. Here we focus on critical quenches with $m_f \equiv m(g_f) \ll m(g_i)\equiv m_i$. 
For $g_{i,f}>1$, in \eref{chi1} $\tan(\Delta \alpha_k) \simeq k/(2m_f)$ for $k \ll m_f$,  
$\simeq 1$ for $m_f \ll k \ll m_i$, whereas $\simeq m_i/(2k)$ for $m_i \ll k $. 
Taking into account \eref{angleappr} one therefore finds 
\be
\chi_1(g_f,g_i) = - 1/(4\pi) \ln(m_i/m_f) + \rm regular \;\;terms.
\label{chi1-leading}
\ee
(If the lattice spacing $a$, presently set to $1$, is reinstated into the calculation, the prefactor of the logarithm becomes $\propto a$ and is therefore not universal.)
The same result is obtained for $g_i>1>g_f$, whereas $\chi_1$ has the opposite sign compared to \eref{chi1-leading} whenever $g_i<1$.
A further derivative with respect to $g_f$ gives the singular behavior of $\chi_2 \simeq 1/(g_f-1)$ in agreement with the scaling law \eref{scaling} in $d=1$ with the proper value $\nu=1$ in two dimensions.

The final goal of our analysis is to calculate the statistics of the work $W$ close to the threshold $\Delta E_0$ and to criticality $g_{i,f}\!\simeq\!g_c$, focussing 
on the $t$-dependent part 
$f_c(g_f,g_i,t)=-1/L\sum_{k>0}\ln\left[ 1+(\tan \Delta \alpha_k)^2 \rme^{-2iE_k(g_f)t} \right]$ 
of the exponent in the characteristic function.
In order to extract its low-$w$ asymptotics we consider as before the critical Casimir free energy
$f_c(g_f,g_i,-iR)$, we extract its asymptotic expansion for large $R$ and finally we continue analytically an expansion of the form of \eref{expansiongenerating}. Since this asymptotics is again determined by the low-momentum
modes it is convenient to take the scaling limit, which gives
\be
\label{scalingcasimir}
f_c=- \int _0^{+\infty}\!\!\frac{\rmd k}{2\pi}\ln\left[ 1+(\tan \Delta\alpha_k)^2\rme^{-2 \sqrt{k^2+m_f^2} R} \right],
\ee
where the $\theta_k$'s in $\Delta\alpha_k$ are given by \eref{angleappr}.
Note  that since our focus is on universal quantities, the expression above can be regularized at large momenta via a convenient cutoff scheme. In particular, we imply that in doing the final analytic continuation $R\rightarrow it+\tau_0$ we 
keep a small real part, which effectively cuts off the contribution of ultraviolet modes in \eref{scalingcasimir}. 

In order to classify the various possible low-$w$ behavior of $P(W)$, we consider
the small-$k$ expansion of $\tan(\Delta\alpha_k)$ in \eref{scalingcasimir}. 
Below we summarize the main results of this analysis, the details of which will be presented elsewhere~\cite{gs-11b}.
When the initial and final states are both either ferromagnetic or paramagnetic, 
the problem can be simplified expanding in the difference $g_f-g_i$, 
and  $\tan(\Delta\alpha_k)\simeq  \pm \rho_- k/(2m_f)$
for $g_{i,f}\!\gtrless\! 1$, where $\rho_\pm \equiv [(m_i\pm m_f)/m_i]^{\mp 1}$. 
Calculating the asymptotics of \eref{scalingcasimir} for $x\!=\!m_fR\!\gg\! 1$ one obtains the scaling function in \eref{scalingfunction} 
\be
F(x\gg 1) \simeq \rho_-^2  \left(- \frac{\rme^{-2x}}{32\sqrt{\pi x}}\right).
\label{eq:Cas-ss}
\ee 
For a two-dimensional classical Ising model in a strip a critical Casimir free energy with the asymptotic dependence on $x=R/\xi_+ \gg 1$ given by \eref{eq:Cas-ss} with $m_f \ll m_i$ corresponds to the case of 
fully magnetized boundaries (effectively realizing the so-called $(+,+)$ boundary conditions \cite{diehl}) and $T<T_c$, where $T$ is the temperature and $T_c$ its critical value in the bulk. By duality, this corresponds also to the case of effectively paramagnetic boundaries (realizing the so-called $(O,O)$ boundary conditions \cite{diehl}) and $T>T_c$ \cite{ES-94,ZR-10}.  
In both these instances the boundaries have the same paramagnetic/ferromagnetic character as the bulk.  As expected, instead, $F$ vanishes for $m_i=m_f$, as no quench has occurred.
After the analytic continuation, the characteristic function can be expanded as in \eref{expansiongenerating} and one obtains for $P(W)$ a peak at the threshold $\Delta E_0$ plus a continuum
starting at $2m$, which describes pairs of quasi-particles. The corresponding edge singularity is universal and fully determined by the exponent of $A(R)$, i.e.,
\be
P(w)\!\propto\! \delta(w)\!+\!\frac{\sqrt{\pi}}{4}\frac{\vartheta(w-2m_f)}{\delta}\rho_-^2  
\sqrt{\frac{w\!-\!2m_f}{m_f}} + \ldots
\ee
where $\delta=4\pi/L$ is the two-particle level spacing, $w\equiv W\!-\!\Delta E_0$, and $\vartheta$ is the unit step function.
Perturbation theory breaks down as soon as either the final state is critical or the quench takes the system across the phase transition. In the first case, if $g_f=1$ (or $g_i=1$),
$\tan(\Delta \alpha_k) \simeq 1$, while 
\be
\label{pole}
\tan(\Delta \alpha_k) \simeq \pm 2 \rho_+ m_f k^{-1}
\ee
for $g_i\!\lessgtr\!1\!\lessgtr\! g_f$
In both cases the behavior of $P$ at small $w$ is significantly different from the one obtained above. Indeed, while for quenches at criticality one obtains a scaling function 
$F(x=0)=-\pi/48$, 
consistent with the results of conformal field theory~\cite{cc-rev}, 
for quenches across the quantum critical point and $x\gg 1$ one has 
\be
F(x)\simeq - \rho_+  x \rme^{-x} + \frac{2}{\sqrt{\pi}}  \rho_+^2 x^{\frac{3}{2}}\rme^{-2x} + \ldots,
\label{eq:Cas-acr}
\ee
where we kept also the leading term in  $\exp(-2x)$, which controls the edge singularity associated to the two-particle continuum. The behavior corresponding to 
\eref{eq:Cas-acr} with $m_f \ll m_i$ is observed in the critical Casimir
effect for $T>T_c$ and $(+,+)$ boundary conditions  or, by duality, for $T<T_c$ and $(O,O)$ boundary conditions, i.e., whenever the boundaries and the bulk have  opposite paramagnetic/ferromagnetic characters. 
Comparing  the two cases corresponding to \erefs{eq:Cas-ss} and \reff{eq:Cas-acr} for $m_f \ll m_i$, we note that the associated results for $F(x)$ are consistent with a change in the effective boundary conditions upon changing the ferromagnetic/paramagnetic character of the initial state~\cite{cg-10,gs-11b}. 
Taking now the analytic continuation of the previous results we obtain 
\be
P(w)\propto \delta(w)+\frac{\pi^2\vartheta(w)}{12\delta}+\ldots,
\ee
for quenches to the quantum critical point,
i.e., the continuum starts exactly at the edge, as expected from having gapless excitations in the final state, while for quenches across the quantum critical point we have
\begin{eqnarray}
\label{across}
P(w)&\propto & \delta(w)+\frac{2\pi}{\delta} (2m_f\rho_+) \delta(w-m_f) \nonumber \\
&+& \frac{2\pi^2}{\delta^2} (2m_f\rho_+)^2\delta(w-2m_f) \nonumber \\
&+&  \frac{\vartheta(w\!\!-\!\!2m)}{\delta} (2\rho_+)^2 \left[ \frac{m_f}{w\!-\!2m_f} \right]^{\frac{3}{2}}\!
\end{eqnarray}
The emergence of the delta-function peaks at $m_f,2m_f$, etc.. is a non-perturbative phenomenon  consequence of the kinematic pole in \eref{pole}. Notice that while the edge singularity at $2m$ appears to be not integrable, the presence of a cutoff at $w-2m_f \approx1/L$ is implied by the finite size of the system.

Summarizing, we established a connection between the statistics of the work done during a quantum quench and the critical Casimir effect. This relation allowed us to demonstrate  how universal features emerge in this non-equilibrium evolution depending on the initial conditions.  We expect that the extension of this analysis to different systems and protocols will help elucidating the role of universality in the dynamics of closed and strongly correlated quantum systems.

\emph{Acknowledgements} -- We are grateful to E. Altman, P. Calabrese,
V. Gritsev, F. Essler, R. Fazio, S. Kehrein, G. Mussardo and A. Polkovnikov for useful discussions. 
AS would like to acknowledge the hospitality of the KITP Santa Barbara. This research was supported in part by the National Science Foundation under Grant No. NSF PHY05-51164.  



\begin{thebibliography}{10}

\bibitem{Gre-Kino} 
M. Greiner {\it et al.}, Nature {\bf 415}, 39
(2002); M. Greiner {\it et al.}, Nature {\bf 419}, 51 (2002);  
T. Kinoshita, T. Wenger, and D. S. Weiss,
Nature {\bf 440}, 900 (2006).

\bibitem{Polkovnikov10} 
A.~Polkovnikov, K.~Sengupta, A.~Silva and M.~Vengalattore, arXiv:1007.5331.  

\bibitem{Zurek} W. H. Zurek, U. Dorner, P. Zoller, Phys. Rev. Lett.
\textbf{95}, 105701 (2005); A. Polkovnikov, Phys. Rev. B \textbf{72},
161201(R) (2005); A. Polkovnikov and V. Gritsev, 
 Nature Phys. {\bf 4}, 477 (2008). 

\bibitem{Calabrese} P. Calabrese and J. Cardy, J. Stat. Mech. (2007)
P10004.

\bibitem{Entropy} A. Polkovnikov, Annals of Phys. \textbf{326}, 486 (2011). 

\bibitem{Silva08} A. Silva, Phys. Rev. Lett. \textbf{101}, 120603 (2008). 

\bibitem{Jarzynski} C. Jarzynski, Phys. Rev. Lett. {\bf 78}, 2690
(1997)

\bibitem{Lutz} J. Kurchan, arXiv:cond-mat/0007360v2; 
P. Talkner, E. Lutz, and P. H\"{a}nggi, Phys. Rev. E {\bf 75}, 050102 (2007).

\bibitem{work} M. Heyl, S. Kehrein, arXiv:1006.3522; A. Faribault, P. Calabrese, J.-S. Caux
 J. Stat. Mech. (2009); P03018; F. N.C. Paraan, A. Silva, Phys. Rev. E \textbf{80}, 061130 (2009).  

\bibitem{fidelity} L. C. Venuti and P. Zanardi, Phys. Rev. Lett. \textbf{99}, 095701
(2007); W.-L. You, Y.-W. Li, S.-J. Gu, Phys. Rev. E  \textbf{76}, 022101
(2007); D. Schwandt, F. Alet, and S. Capponi, Phys. Rev. Lett.
 \textbf{103}, 170501 (2009); A. F. Albuquerque, {\it et al.}, Phys. Rev. B  \textbf{81}, 064418 (2010); C. De Grandi, V. Gritsev, and A. Polkovnikov
Phys. Rev. B  \textbf{81}, 012303 (2010).  

\bibitem{cc-rev}
M. Krech, \emph{Casimir Effect in Critical Systems} (World Scientific,
Singapore, 1994); 
A. Gambassi, 
J. Phys.: Conf. Ser. {\bf 161}, 012037 (2009).

\bibitem{diehl} 
H. W. Diehl, 
in {\it Phase Transitions and Critical Phenomena} vol 10 ed C. Domb and J. L.
Lebowitz (Academic, 1986); 
Int. J. Mod. Phys. B {\bf 11}, 3503 (1997). 

\bibitem{Has-10}
M. Hasenbusch, Phys. Rev. B {\bf 82}, 104425 (2010).

\bibitem{ELoschmidt}
R. A. Jalabert and H. M. Pastawski, Phys. Rev. Lett. {\bf 86}, 2490
(2001); Z. P. Karkuszewski, C. Jarzynski, and W. Zurek, Phys. Rev.
Lett. {\bf 89}, 170405 (2002); H. T. Quan, \it et al. \rm
Phys. Rev. Lett. {\bf 96}, 140604 (2006); D. Rossini, \it et al. \rm
Phys. Rev. A {\bf 75}, 032333 (2007).



\bibitem{Sachdev} S. Sachdev, {\it Quantum Phase Transitions}
(Cambridge University Press, Cambridge, 1999).

\bibitem{note2} While an additional, parity-sector changing, multiplicative factor of the form $(1+v_0\gamma^{\dagger}_0)$  is admissible on the rhs of this equation, it will be omitted here. Ultimately, this zero-momentum mode will emerge by taking appropriately the thermodynamic limit (see Eq.(\ref{across})).

\bibitem{note} Particular care has to be taken here. Indeed, expanding $\tan(2\theta_k)\simeq k/(g-1+k^2/2)$,
a finite momentum scale $k^*=\sqrt{2 \Delta}$ controlling the asymptotic behavior of $\theta_k$ naturally emerges. 
%
However if one reinstates the energy scale $J$ and the lattice spacing $a$, then
$k^*=\sqrt{2\Delta/(ca)} $,
where $c=Ja$ and $\Delta=J| g-1 |$ in dimensionfull units. Since the scaling limit corresponds to taking $a \rightarrow 0$, $J \rightarrow +\infty$ and $g \rightarrow 1$ in such a way as to keep $\Delta$ and $c$ finite then $k^* \rightarrow +\infty$.

\bibitem{gs-11b}
A.~Gambassi and A.~Silva, to appear (2011).

\bibitem{ES-94}
R.~Evans and J.~Stecki, Phys. Rev. B {\bf 49}, 8842 (1994).

\bibitem{ZR-10}
J.~Rudnick, {it et al.}, 
Phys. Rev. E {\bf 82}, 041118 (2010).

\bibitem{cg-10}
A.~Gambassi and P.~Calabrese, arXiv:1012.5294 (2010)

\end{thebibliography}
\end{document}